\documentclass[aps,english,twocolumn,superscriptaddress]{revtex4}
\usepackage{graphicx}
\usepackage{amssymb}
\usepackage{babel}

\begin{document}

\title{Static charging of graphene and graphite slabs}

\author{M. Topsakal}
\affiliation{UNAM-Institute of Materials Science and
Nanotechnology, Bilkent University, Ankara 06800, Turkey}
\author{S. Ciraci}
\affiliation{UNAM-Institute of Materials Science and
Nanotechnology, Bilkent University, Ankara 06800, Turkey}
\affiliation{Department of Physics, Bilkent University, Ankara
06800, Turkey}

\date{\today}

\begin{abstract}
The effect of external static charging of graphene and its flakes are investigated by using first-principles calculations. While the Fermi level of negatively charged graphene rises and then is quickly pinned by the parabolic, nearly free electron like bands, it moves down readily by removal of electrons from graphene. Excess charges accumulate mainly at both
surfaces of graphite slab. Even more remarkable is that
Coulomb repulsion exfoliates the graphene layers from both
surfaces of positively charged graphite slab. The energy level
structure, binding energy and and spin-polarization of specific
adatoms adsorbed to a graphene flake can be monitored by charging.
\end{abstract}

\maketitle

Graphene\cite{gra1} is a semimetal having 
conduction and valance bands which cross linearly at the Fermi level ($E_F$).
The resulting electron-hole symmetry reveals itself in an
ambipolar electric field effect, whereby under bias voltage the charge
carriers can be tuned continuously between electrons and holes in
significant concentrations. Excess electrons and holes can be also
achieved through doping with foreign
atoms.\cite{doping1,doping2,doping3} For example, adsorbed alkali
atoms tend to donate their valence electrons to $\pi^*$-bands of
graphene. The excess electrons results in the metalization of
graphene.\cite{can} Hole doping is achieved by the adsorption of
bismuth or antimony.\cite{gierz} However, the system remains 
electrically neutral through either way of doping. Recently, carrier
concentration and spatial distribution of charge are also changed
for very short time intervals by photoexcitation of electrons from the
filled states leading to the photoexfoliation of
graphite.\cite{zewail,tomanek,tomanek2}

In this letter, we demonstrate that the properties of graphene can be
 modified either by direct electron injection into it
or electron removal from it; namely by charging the system
externally. Remarkably, the Coulomb repulsion exfoliates the
graphene layers from both surfaces of positively charged graphene
slab. This result may be exploited to develop a method for intact
exfoliation of graphene. In addition to exfoliation, the energy level
structure, density of states,\cite{cohen} binding energies and desorption of specific adatoms can
be monitored by charging.

\begin{figure}
\includegraphics[width=8cm]{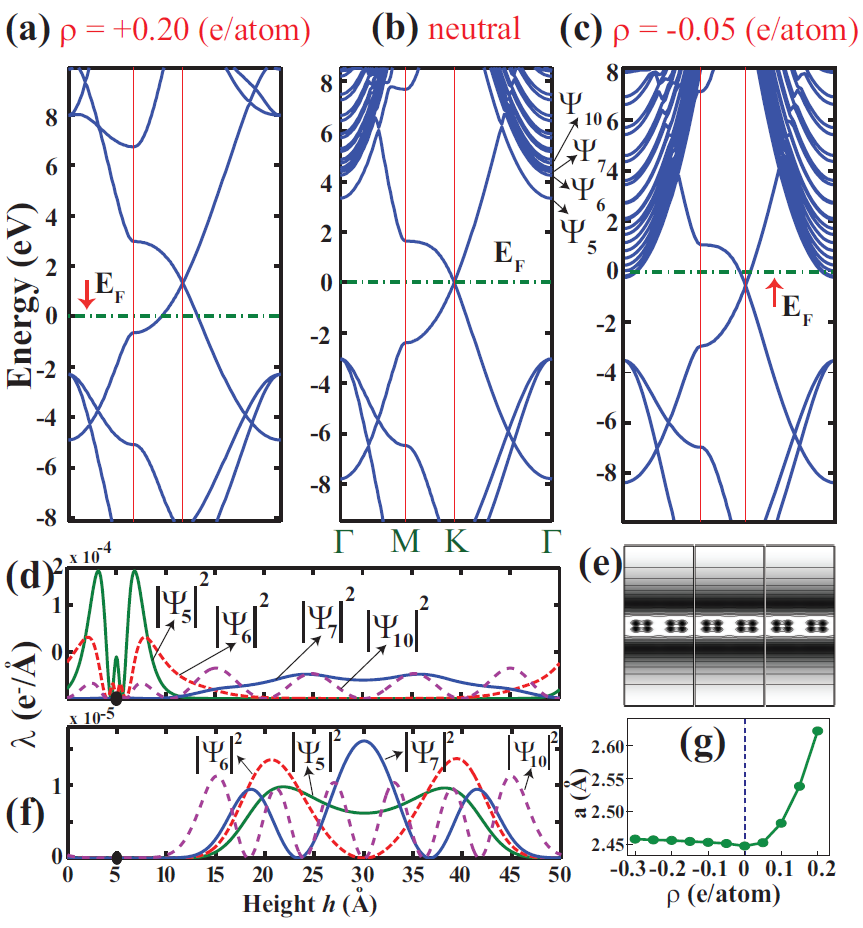}
\caption{(Color Online)  Energy band structure of charged and
neutral graphenes. (a) Positively charged graphene by $\rho$=+0.20
e/atom. (b) Neutral. (c) Negatively charged graphene by
$\rho$=-0.05 e/atom, where excess electrons start to occupy the
surface states. Zero of energy is set to Fermi level. (d) Planarly
averaged charge density ($\lambda$) of states, $\Psi_{5-10}$, of neutral
graphene. (e) Charge contour plots of the lowest surface state,
$\Psi_{5}=\Psi_{S}$ in a plane perpendicular to graphene. (f) Same
as (d) after charging with $\rho$=-0.05 e/atom. (g) Variation of
lattice constant $a$ of graphene as a function of charging.}

\label{fig:Figure-graphene}
\end{figure}

Our results are predicted through first-principles plane wave
calculations carried out within density functional theory (DFT)
using  projector-augmented wave potentials.\cite{paw} The exchange
correlation potential is approximated by local-density
approximation (LDA). We also performed GGA+vdW (generalized gradient 
approximation including van der Waals 
corrections\cite{grimme}) for a better account of VdW interlayer interactions
between graphite slabs. 
A plane-wave basis set with kinetic energy
cutoff of 500 eV is used. All atomic positions and lattice
constants are optimized by using the conjugate gradient method,
where the total energy and atomic forces are minimized. The
convergence for energy is chosen as 10$^{-5}$ eV between two
steps, and the maximum force allowed on each atom is less than 0.01 eV/\AA{}.
 The Brillouin zone (BZ) is sampled by (15x15x5) special
\textbf{k}-points for primitive unit cell. Calculations for neutral, as well as
charged systems are carried out by using VASP
package.\cite{vasp} Two-dimensional graphene is treated within
periodic boundary conditions using the supercell method having more than 50
\AA{} separation between adjacent layers. The amount of charging,
$\rho$, is specified as either positive charging, i.e. electron depletion ($\rho
> 0$), or negative charging, i.e. excess
electrons, in units of $\pm$ electron (e) per carbon atom or per
unit cell. For charged calculations, additional neutralizing background charge
is applied.\cite{payne}

The work function of neutral graphene is calculated to be 4.77 eV.
Lowest two parabolic bands $\Psi_5$ and $\Psi_6$ in
Fig.~\ref{fig:Figure-graphene}(b) have effective masses m$^*$=1.05
and 1.02 m$_e$ (free electron mass) in the $xy$-plane parallel to the atomic
plane of graphene. 
Hence they are nearly free electron like
(NFE) in 2D , but they are bound above the graphene plane. As
shown, in Fig.~\ref{fig:Figure-graphene}(d) and (f), these
"surface" states\cite{posternak} can be expressed as $\Psi_{S}
\sim e^{i\bf{k_{\parallel}.r_{\parallel}}} \Phi (z)$, where
$\bf{r_{\parallel}}$ and $\bf{k_{\parallel}}$ are in the
$xy$-plane. Parabolic bands at higher energies becomes NFE in 3D. When the
electrons are removed, the Fermi level is lowered from the Dirac
point and positively charged graphene attains metallic behavior as
in Fig.~\ref{fig:Figure-graphene}(a). At the end, the work
function increases. However, under negative charging, whereby
electrons are injected to the graphene, Fermi level raises above
the Dirac point and eventually becomes pinned by NFE parabolic
bands as in Fig.~\ref{fig:Figure-graphene}(c). These parabolic NFE
bands start to get occupied around $\rho$=-0.015 e/atom (or
surface excess charge density $\sigma$=-0.0926 C/m$^{2}$). Upon
charging the bound charge of $\Psi_{S}$ states are further removed
from graphene as shown in Fig.~\ref{fig:Figure-graphene}(f). This
situation can be interpreted as the excess electrons start to
spill out towards vacuum. Figure~\ref{fig:Figure-graphene}(g) shows
another important effect of charging where the lattice
constants increase with positive charging. On the other hand, negative charging
has little effect on lattice constants, since the excess electrons mostly
spill out.

\begin{figure}
\includegraphics[width=7.5cm]{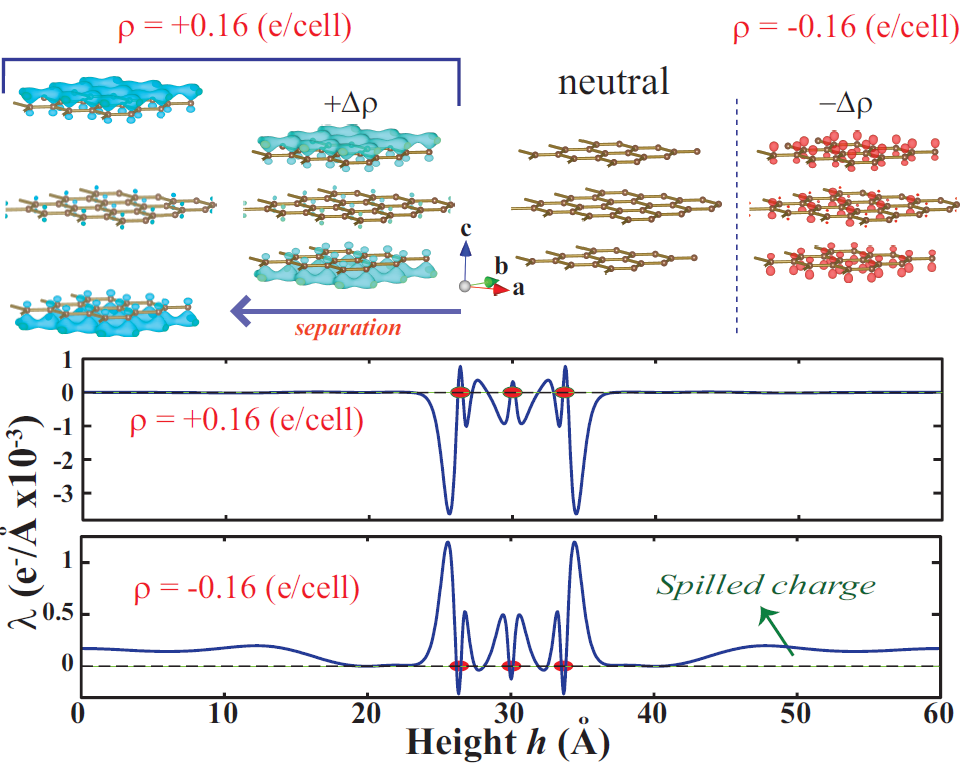}
\caption{(Color Online) Exfoliation of graphene layers from both surfaces of a
3-layer graphite slab (in AB-stacking) caused by electron removal.
Isosurfaces of difference charge density, $\Delta \rho$, show the
electron depletion. The excess charge on the negatively charged
slab is not sufficient for exfoliation. The distributions of
planar averaged charge density ($\lambda$) perpendicular to the graphene plane
are
shown below both for positive and negative charging (calculations are performed
by GGA+vdW).}

\label{fig:Figure-3layer}
\end{figure}

The effect of charging on a graphite slab consisting of 3 layers of
graphene is better seen in
Fig.~\ref{fig:Figure-3layer}. When negatively charged, the excess electrons
are mainly accumulated on both surfaces, but with smaller amount
at the middle layer. The effect of charging on structure is
minute, since the bonds are intact and the excess electrons
rapidly spill out towards the vacuum. However, the situation is dramatically
different for the case of positive charging. The charge
isosurfaces in Fig.~\ref{fig:Figure-3layer} shows that positive
charge, occurs mainly  on both surfaces (i.e.
first and third graphene layers), whereas the middle graphene has
relatively small positive charge. This is an expected result for a
metallic system. {The interlayer interaction in the neutral
3-layer slab is attractive and is calculated to be 17 (36)
meV/atom calculated by LDA (GGA+vdW), which becomes even weaker upon
depopulation of $\pi$-orbitals. GGA+vdW calculations predict that a threshold
charge, Q=0.16 e/cell gives rise to exfoliation of two outermost layers. LDA
calculations yield 
relatively lower threshold charge of Q=0.14 e/cell. We also performed a
systematic analysis
of exfoliation for thicker slabs consisting of 5-10 layers of graphene. We found
that 
the threshold charge increases with increasing slab thickness. However, our
analysis based on the planar averaged charge densities suggest that the
exfoliation of outermost layers occurs when approximately the same amount of
positive charge is accumulated on the outermost layers. For example, the
exfoliation of 3-layer and 6-layer graphene flakes take place when their
outermost layers have positive charge of 0.065 and 0.066 e/cell, respectively.
On the other hand, increasing of threshold charge by going from 3-layer to
6-layer occurs due to the charge spill to the inner layers. This situation can
be explained by a simple electrostatic model, where the outermost layers of
slabs is modeled by uniformly charged planes, which yield repulsive interaction
independent of their separation distance, i.e. $F \propto
q^{2}/(A\cdotp\epsilon_{0})$, where q is excess positive charge per unit cell
with the area A. Nonetheless, these values of
charging are quite high, and can be attained in small flakes locally
by the tip of Scanning Tunnelling Microscope.

\begin{figure*}
\includegraphics[width=12cm]{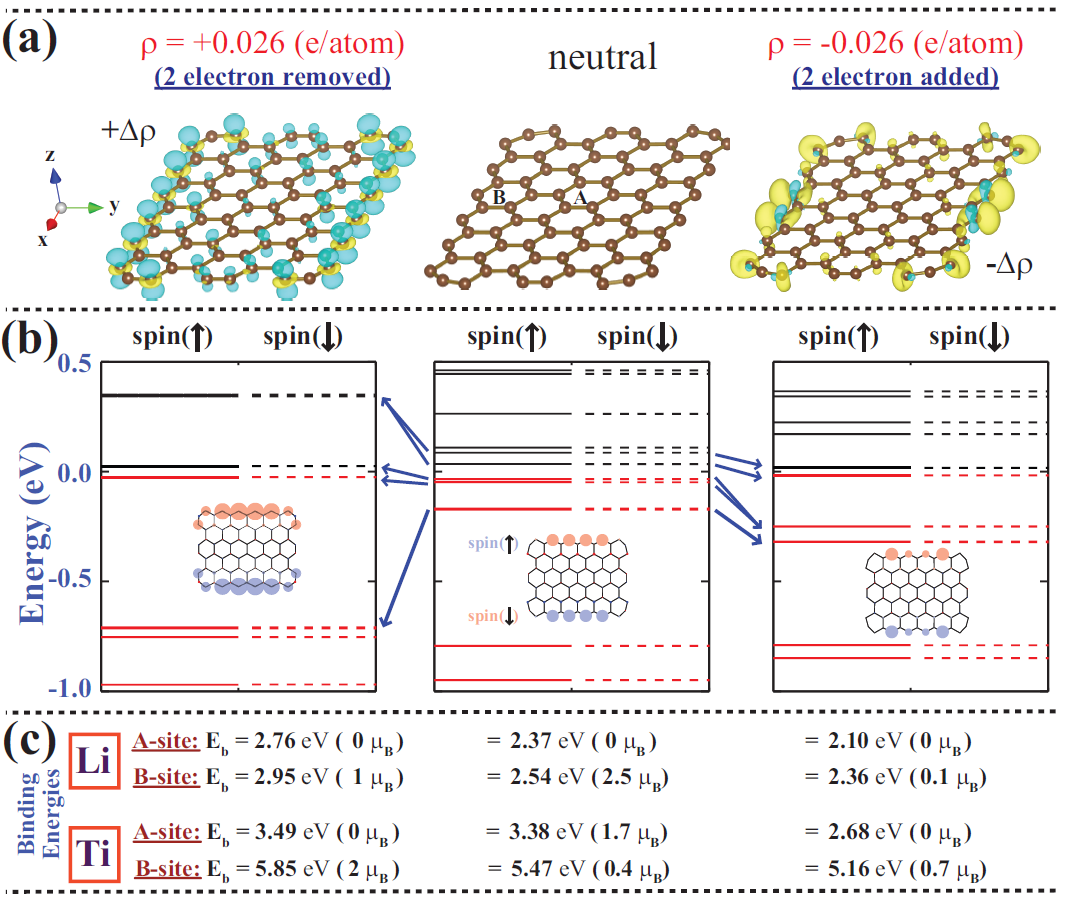}
\caption{(Color Online) Effect of charging on graphene flake
consisting of 78 carbon atoms. (a) Isosurfaces of difference
charge density $\Delta \rho$ of positively charged, neutral and
negatively charged slabs. (b) Corresponding spin-polarized energy
level structure. Solid and continuous levels show spin up and spin
down states. Red and black levels indicate filled and empty
states, respectively. Distribution of magnetic moments at the
zigzag edges are shown by insets. Zero of energy is set to Fermi
level.(c) Variation of binding energy and net magnetic moment of
specific adatoms adsorbed in two different positions, namely
A-site and B-site indicated in (a).}

\label{fig:Figure-flake}
\end{figure*}

Ultra-fast graphene ablation was directly observed by means of
electron crystallography.\cite{zewail} Carriers excited by
ultra-short laser pulse transfer energy to strongly coupled optical
phonons. Graphite undergoes a contraction, which is subsequently
followed by an expansion leading eventually to laser-driven
ablation.\cite{zewail} Much recently, the understanding of
photoexfoliation have been proposed, where exposure to femtosecond
laser pulses has led to athermal exfoliation of intact
graphenes.\cite{tomanek} Based on time dependent DFT calculations
(TD-DFT), it is proposed that the femtosecond laser pulse rapidly
generates hot electron gas at $\sim20.000$ K, while graphene layers are
vibrationally cold. The hot electrons are spill out, leaving
behind a positively charged graphite slab. The charge deficiency
accumulated at the top and bottom surfaces lead to athermal
excitation.\cite{tomanek} The exfoliation in static charging
described in Fig.~\ref{fig:Figure-3layer} is in compliance with
the understanding of photoexcitation revealed from TD-DFT
calculations, since the driving force which leads to the
separation of graphenes from graphite is related mainly with
electrostatic effects in both process.

The effects of charging becomes emphasized when the size of
graphene flake is small. In this respect, the flake behave like a
quantum dot and hence energy level structure is affected strongly.
The flake we consider has a rectangular shape and hence it
consists of armchair, as well as zigzag edges as shown in
Fig.~\ref{fig:Figure-flake}. It is therefore antiferromagnetic
ground state when neutral. Isosurfaces of difference charge
density, $\Delta \rho$, of the same flake for three different
charge state are shown in Fig.~\ref{fig:Figure-flake}(a). $\Delta
\rho$ is calculated by subtracting the total charge density of the
neutral flake from that of charged ones. For a better comparison,
charge density of the neutral flake is calculated using the atomic
structure of the charged ones. It is seen that the edge states due
to zigzag edges are most affected from charging. In
Fig.~\ref{fig:Figure-flake} (a), while charge is depleted mainly
from edge states, excess electrons are accumulated predominantly
at the zigzag edges. As shown in Fig.~\ref{fig:Figure-flake} (b),
while the antiferromagnetic state of the flake is unaltered,
charging causes emptying and filling of HOMO and LUMO states,
changing of level spacings and their energies relative to vacuum
level. Additionally, magnetic moments of zigzag edge atoms are
strongly affected depending on the sign of charging in
Fig.~\ref{fig:Figure-flake} (b). In particular, the binding
energies and magnetic moments of specific adatoms depends on its
position and charging of the flake. In Fig.~\ref{fig:Figure-flake} (c) we consider Li and Ti,
which normally adsorbed to graphene by donating charge. Generally,
the binding energies increases (decreases) with positive
(negative) charging. We also found that the effects of monopole and dipole corrections 
on the effects of charging on the binding energies is minute. For example, the binding 
energy of Li, when 2 electrons are removed, increases from 2.756 to 2.764 upon
corrections. However, the effect of 
charging becomes more pronounced when the adatom is placed close to the edge of
positively charged flake since the additional charges are
mostly confined at the edges. Similarly, the magnetic moment at the
adatom site varies depending the adsorption site and charging
state of the flake.

In summary, we revealed the dramatic effects of static and external
charging of graphene and its flake. Charging through electron
depletion of graphite surfaces leads to exfoliation of graphene.
We also show that the binding energy and local magnetic moments of
specific adatoms can be tuned by charging.

We thank the DEISA Consortium (www.deisa.eu), funded through the
EU FP7 project RI-222919, for support within the DEISA Extreme
Computing Initiative.  We acknowledges partial financial support
from The Academy of Science of Turkey (TUBA).


\begin{thebibliography}{15}

\bibitem{gra1}
K. S.  Novoselov,  A. K.  Geim,  S. V.  Morozov, D.  Jiang,  Y.
Zhang, S. V.  Dubonos, I. V. Grigorieva, and A. A. Firsov, Science
\textbf{306}, 666 (2004).


\bibitem{doping1} 
X. Wang, X. Li, L. Zhang,  Y. Yoon, P. K. Weber, H. Wang, J. Guo, H. Dai, Science
\textbf{324}, 768 (2009).

\bibitem{doping2} 
T. O. Wehling, K. S. Novoselov, S. V. Morozov, E. E. Vdovin, M. I.
Katsnelson, A. K. Geim, and A. I. Lichtenstein, Nano Lett.
\textbf{8}, 173 (2008).

\bibitem{doping3}
H. Sevincli, M. Topsakal, E. Durgun and S. Ciraci, Phys. Rev. B,
\textbf{77}, 195434 (2008).

\bibitem{can}
C. Ataca, E. Akt\"{u}rk, S. Ciraci, and H. Ustunel,
Appl. Phys. Lett. \textbf{93}, 043123 (2008).

\bibitem{gierz} 
I. Gierz, C. Riedl, U. Starke, C. R. Ast, and K. Kern, Nano Lett. \textbf{8}, 4603 (2008).

\bibitem{zewail}
F. Carbone, P. Baum, P. Rudolf, and A. H. Zewail,
Phys. Rev. Lett. \textbf{100}, 035501 (2008).

\bibitem{tomanek}
Y. Miyamoto, H. Zhang, and D. Tomanek,
Phys. Rev. Lett. \textbf{104}, 208302 (2010).

\bibitem{tomanek2}
R. K. Raman, Y. Murooka, C.-Y. Ruan, T. Yang, S. Berber, and D. Tomanek,
Phys. Rev. Lett. \textbf{101}, 077401 (2008).

\bibitem{cohen}
K. T. Chan, H. Lee, and M. L. Cohen, Phys. Rev. B \textbf{83}, 035405 (2011).

\bibitem{paw}
P. E. Blochl, Phys. Rev. B \textbf{50}, 17953 (1994).

\bibitem{grimme}
S. Grimme, J. Comp. Chem. \textbf{27}, 1787 (2006).

\bibitem{vasp}
G. Kresse, J. Furthmuller, Phys. Rev. B \textbf{54}, 11169 (1996).

\bibitem{payne}
G. Makov and M. C. Payne, Phys. Rev. B \textbf{51}, 4014 (1995).

\bibitem{posternak}
M. Posternak, A. Baldereschi, A.J. Freeman and E. Wimmer, Phys.
Rev. Lett. \textbf{52}, 863 (1984).

\end{thebibliography}
\end{document}